\documentstyle[aps,pre,twocolumn,epsfig]{revtex}

\begin{document} 

\title{Pattern Formation of Ion Channels with State Dependent
  Electrophoretic Charges and Diffusion Constants in Fluid  Membranes}  
\author{S. C. Kramer and R. Kree} 
\address{Institut f\"ur Theoretische
  Physik, Georg-August-Universit\"at G\"ottingen, Bunsenstr. 9, 37073
  G\"ottingen} 
\author{\begin{center}\parbox[b]{0.75\linewidth}{ A
      model of mobile, charged ion channels in a fluid membrane is
      studied. The channels may switch between an open and a closed
      state according to a simple two-state kinetics with constant
      rates. The effective electrophoretic charge and the diffusion
      constant of the channels may be different in the closed and in
      the open state. The system is modeled by densities of channel
      species, obeying simple equations of electro-diffusion. The
      lateral transmembrane voltage profile is determined from a
      cable-type equation. Bifurcations from the homogeneous,
      stationary state appear as hard-mode, soft-mode or hard-mode
      oscillatory transitions within physiologically reasonable ranges
      of model parameters. We study the dynamics beyond linear
      stability analysis and derive non-linear evolution equations
      near the transitions to stationary patterns.}\end{center}
}



\maketitle




\section{Introduction}
\label{sec:intro}

Spontaneous pattern formation of ion channels in cell membranes has
been a subject of continuous interest during recent years
\cite{fromherz1988a,fromherz1995a,jaffe1977}.
Spatially modulated distributions of ion channels or pumps are
ubiquitous in many cells. They are closely related to important
biological functions like some early stages of morphogenesis
\cite{harold1990} or the storage and processing of information in
neural tissue \cite{scott1975}. Many physiological regulation processes
involve changes of ion fluxes through cell membranes
on time scales ranging from milliseconds to hours \cite{kandel1985}.

There have been attempts to explain the aggregation of channel
proteins in a fluid membrane on the basis of
thermodynamic models \cite{fromherz1988a}, \cite{savtchenco1994} and
references therein.
Typically the system of channels is driven out of equilibrium due to
ionic concentration gradients and transmembrane fluxes. Therefore, a
number of authors have put forward models for spontaneous pattern
formation based upon systems of semi-phenomenological, non-equilibrium
equations of motion for the densities of channel proteins, the charge
densities of ions and the intra- and extracellular voltage
\cite{fromherz1988a,jaffe1977,Leonetti1997}. In many important systems
like axons or cells in neural or muscular tissue, there is a bulk
volume of aqueous solution of ions on one side of the membrane whereas
on the other side, there is only a thin layer of electrolyte, whose
thickness usually is in the range of $10\cdots 100$ nm. See
Fig.\ref{fig:fig1} For a schematic picture of the situation. Under
these conditions, the models may be simplified further. Ion densities
may be eliminated approximately, leading to a coupling of the
densities of channels to the lateral transmembrane voltage profile
within that layer. The voltage profile has to be determined from a
quasi one-dimensional or two-dimensional cable equation
\cite{fromherz1995a,Leonetti1998,Leonetti1998b}. This contains
transmembrane currents and thus channels are effectively coupled via
the transmembrane voltage.

In the present work we will follow this approach and analyze a model
of charged ion channels, which may switch between a closed and an open
state, thereby also changing their effective electrophoretic charge
and their diffusion constant. The effective electrophoretic charge can
be of both signs - although the proteins are usually only negatively
charged - if the electro-osmotic effect is taken into account
\cite{Leonetti1997}. 
 
We consider this model as a natural, minimal model of mobile channels
with kinetics of internal states. As the conformations, which change
the states of an ion channel \cite{aidley1996,hille1992}, may involve charge
transport across the membrane \cite{hille1992,lehninger2000} as well
as binding to other molecules, for instance phosphorylation via protein
kinase or immobilization due to binding to the cytoskeleton
\cite{lehninger2000,peng1986}, the state dependence of charge and
diffusion constant should be taken into account.

In \cite{fromherz1995a} a special case of the present model has
been studied already. The authors of this work considered a reservoir
of closed channels in the well-stirred approximation without coupling
to the lateral voltage profile. We will see below that our model leads
to qualitatively different behavior in a wide range of parameters but
still contains the results of \cite{fromherz1995a}. The
necessary limit is that closed channels diffuse faster than the open
ones and that their electrophoretic charge vanishes. The present model
is also related to \cite{fromherz1988c}, where a mixture of two
different channels has been considered without a kinetic, which allows
for transitions between the different species.

The basic mechanism, which drives spontaneous pattern formation is
very simple in models, which only involve mobile, charged ion channels
and the lateral profile of the transmembrane voltage
\cite{jaffe1977}. It is based on a feedback loop during which the
channel proteins drift due to lateral voltage gradients and at the
same time modify the voltage gradients by Ohmic voltage drops, caused
by transmembrane currents. Therefore, for example, negatively charged
channels, which cause currents directed into the thin layer of
electrolyte have a tendency to aggregate. Diffusion will counteract
this tendency and thence a variety of bifurcations of the homogeneous
state to an inhomogeneous state becomes possible. We will show below
that our model implies a number of different scenarios for the first
bifurcation from the homogeneous state. 

Depending on parameter ranges either soft-mode instabilities arise,
where periodic patterns appear with a wavenumber $k_c$, which approaches zero
at the transition.  Or there will be different kinds of hard-mode
instabilities, where patterns with $k_c\neq 0$ at the
transition and spatio-temporal patterns emerge with non-vanishing $k_c$ and
non-vanishing frequency $\Omega_c$.

After introducing the model in the next section, we will give results
of the linear stability analysis together with results of non-linear
behavior close to transitions into stationary patterns in subsequent
sections.  For soft-mode instabilities, we derive a Cahn-Hilliard type
equation of motion for the slow modes near the transition. For
hard-mode instabilities, an amplitude equation is derived, which
allows us to separate parameter regions of forward and backward
bifurcation or continuous and discontinuous transitions respectively.

%

\section{The Model}
\label{sec:model}

We consider ion channels moving within a fluid membrane of size $L_x
\times L_y$ which separates a thin layer of electrolyte of width $d$
from an electrolytic bath, as it is sketched in Fig.\ref{fig:fig1}.
Positions ${\mathbf r}$ within the membrane are described by internal
rectilinear coordinates ${\mathbf r}=(x,y)$, $0\le x < L_x$, $0 \le
y <L_y$. If we choose periodic boundary conditions in the $y$
coordinate, the membrane corresponds to a cylindrical cable and is
reminiscent of axonic or dendritic structures for $L_x >> L_y$. Note
that we treat $L_y$ and $d$ as independent quantities, so that the
actual structure in cylindrical geometry looks like on the right hand
side of Fig.\ref{fig:fig1}, with a sub-membrane layer of width~$d$ and
a 'core' of radius $(L_y/2\pi)-d$, which is decoupled from the
sub-membrane layer. For $d=L_y/2\pi$ the sub-membrane layer fills the
entire interior of the cable. In the following we will not distinguish
between these different geometries, as the present work only deals with
infinitely extended systems and hence effects of specific boundary
conditions need not be taken into account.

Ion channels may switch between an open state ('{\em o}') and a
closed state ('{\em c}') according to a simple mono-molecular
chemical reaction scheme
\begin{equation}
  \label{eq:dyn}
  c {\small \begin{array}{c}
\gamma_+ \\ 
\rightleftharpoons \\ 
\gamma_-\\
  \end{array}} o
\end{equation}
with rates $\gamma_-$ and $\gamma_+$. We are interested in a wide
range of kinetic rates, from $10^3/{\rm sec}$ for voltage or ligand gated
channels down to less than one per hour as it is observed for hormone regulated
phosphorylation and dephosphorylation of channel proteins.
The distribution of channels is described by smooth densities
$n_r({\mathbf r},t)$, $r\in \{o,\, c \}$, which obey the equations
\begin{equation}
  \label{eq:rdifg}
  \partial_{t} n_r({\mathbf r}, t) + \nabla {\mathbf j}_r({\mathbf r},t)
=
\sigma_r\left\lbrack \gamma_+
n_c({\mathbf r}, t) - \gamma_-
n_o({\mathbf r}, t)\right\rbrack 
\end{equation}
with $\sigma_r=+1$ for $r=${\em o} and $\sigma_r=-1$ for $r=${\em c}.
The {\em rhs} of this equation takes into account the reaction
kinetics of~(\ref{eq:dyn}). Current densities ${\mathbf j}_r$ of
the channels are assumed to be of the Nernst-Planck form
\begin{equation}
  \label{eq:currentdensities}
  {\mathbf j}_r({\mathbf r},t)   = 
\ -\/ D_r\/\{\nabla n_r({\mathbf r},t)
 -
\beta\/ n_r({\mathbf r},t)\cdot q_r\/{\mathbf E}({\mathbf r},t)\} 
\end{equation}
where $D_r$ and $q_r$ refer to the constants of lateral diffusion and
to the effective electrophoretic charges of open and closed channels,
respectively. The coefficient $\beta$ is the inverse temperature and
$\mathbf{E}({\mathbf r})=-\nabla\/V({\mathbf r})$ denotes the lateral
electric field. Note that we explicitly allow for
different diffusion constants and different electrophoretic charges in
the states {\em o} and {\em c}. The simple form of the current
densities given in (\ref{eq:currentdensities}) may be justified for
rigid protein structures with charged protuberances extending in the
intra- and extracellular electrolyte. As~discussed in~\cite{Leonetti1997}, the effective electrophoretic charges may
be of both signs due to electro-osmotic effects, although proteins are usually negatively charged.

As has been discussed in the literature \cite{Leonetti1998b}, the
Kelvin cable equation or its two dimensional analogue may be used to
calculate the lateral variations of the transmembrane potential
provided that spatial variations of ion concentrations have negligible
effects and that the characteristic length scales of lateral patterns
are large compared to the width of the thin layer of electrolyte. In \cite{Leonetti1998b}, the cable equation is derived by integrating
the Nernst-Planck equation of ion densities together with the Poisson
equation using appropriate boundary conditions at the membrane.
In the following we use the 2-dimensional cable equation \cite{fromherz1995a}
\begin{eqnarray}
  \label{eq:kabelj}
  C_m\partial_t V({\mathbf r},t) &=& \frac{d}{\rho_e}\nabla^{2}V({\bf
    r},t)-G\, V({\bf r},t) \nonumber \\
\nonumber \\
 &&- \lambda n_o({\bf r}, t)\{V({\bf r},t)-E\} \,. 
\end{eqnarray}
$C_m$ denotes membrane capacitance per area, $\rho_e$ 
the resistivity of the electrolyte within the thin layer of width $d$,  
and $\lambda n_o$ is the conductance of open channels. 
We have included a passive, homogeneously distributed transmembrane
conductance $G$. $E$ is the reversal potential 
of ion fluxes through the open channels, which drives the system out
of equilibrium.

We are interested in the stability of the homogeneous and stationary 
solution $\bar{V}, \bar{n}_o, \bar{n}_c$ of (\ref{eq:kabelj})
\begin{eqnarray}
\label{eq:stationary}
\bar{V} &=& \frac{\lambda \bar{n}_o E}{G+\lambda \bar{n}_o}\,,
\nonumber\\
& &  \\
\bar{n}_o = \frac{\gamma_+}{\Gamma}\bar{n} \,, & &\qquad \bar{n}_c
= \frac{\gamma_-}{\Gamma}\bar{n} \,. \nonumber
\end{eqnarray}    
We consider the total density of channels $\bar{n}$ as constant and
define $\Gamma:=\gamma_+ + \gamma_-$. Furthermore, we introduce
appropriately normalized deviations from the stationary and
homogeneous solution as the new fields
$\rho=(n_o-\bar{n}_o)/\bar{n}_o, \ \zeta=(n_c-\bar{n}_c)/\bar{n}_c$
for the open and closed channels, respectively, and $\Phi=\beta
q_o(V-\bar{V})/\bar{V}$ for the membrane potential. Next, we
change to dimensionless lengths and times by using as units the
typical decay length $
\ell_V=\lbrack(\rho_e/d)(\lambda\bar{n}_o+G)\rbrack^{-1/2}$ of $V$ and
a typical diffusion time $\tau=\ell_V^2/D_o$ over this length.  Then
  (\ref{eq:rdifg}), (\ref{eq:currentdensities}) and
(\ref{eq:kabelj}) become dimensionless
\begin{eqnarray}
\label{eq:ma1}
 \partial_t \varrho &=& \nabla\left\lbrace\nabla \varrho + (\varrho+1)\nabla\Phi \right\rbrace
 -\gamma_-(\varrho -\zeta)\,,\\
 \nonumber \\
 \label{eq:ma2}
 \partial_t \zeta &=&D\nabla\left\lbrace\nabla \zeta + q(\zeta+1)\nabla\Phi \right\rbrace
 +\gamma_+(\varrho-\zeta)\,,\\
 \nonumber \\
 \label{eq:ma3}
 \epsilon\,\partial_t \Phi &=& \left\lbrack\nabla^2
   -1\right\rbrack\Phi -\alpha\Phi\varrho - \eta\varrho \,. 
\end{eqnarray}
In these equations we use the ratio of diffusion constants
$D=D_c/D_o$, of electrophoretic charges $q=q_c/q_o$, the relative
conductivity of the membrane
$\alpha=\lambda\bar{n}_o/(\lambda\bar{n}_o+G)$ and a rescaled reversal potential
$\eta=-\alpha(1-\alpha)\beta q_oE$. The parameter
$\epsilon=C_m\rho_eD_o/d$ compares the electrical relaxation (RC) time
per area to the diffusion constant. Note that $\gamma_+$ and
$\gamma_-$ have also been made dimensionless $\gamma_{\pm} \to
\gamma_{\pm}\tau$ by multiplication with $\tau$.

For cell membranes, the parameters in (\ref{eq:rdifg}),
(\ref{eq:currentdensities}) and (\ref{eq:kabelj}) imply time-scales,
which differ by several orders of magnitude. Typical parameter values
are \cite{hille1992} $C_m=1\mu{\rm F/m}^2$, $\rho_e \sim 1\Omega {\rm m}$, $d \sim
10{\rm nm}$, and
$D_r \sim 0.1(\mu \rm m)^2/\rm sec$ for diffusion of mobile proteins. Channel
conductances $\lambda$ are in the $10^{-12}\Omega^{-1}$ range, average
channel densities are around~$10\cdots50\,~(\mu \rm m)^{-2}$ and
electrotonic lengths $\ell_V$ are of the order of a few $\mu\rm m$. Then
the unit $\tau \sim 10^2 \rm sec$, whereas the electrical relaxation time
is below $1 \mu \rm sec$ and it is always the smallest time scale of the
system, even compared to fast kinetics of voltage or ligand gated
channels with time constants in the msec range. The constant
$\epsilon \sim 10^{-7}$. This separation of time scales justifies the
use of the quasi-stationary approximation in (\ref{eq:ma3}) and hence we
will put the {\em rhs} of (\ref{eq:ma3}) equal to zero in the
following. For voltage or ligand gated channels, the time-scale set
by the rate $\Gamma$ is also very fast compared to $\tau$, $\Gamma\sim
10^5$, whereas slower regulation processes have time-scales comparable
to or larger than $\tau$. The limit $\Gamma\gg 1$ will be referred to
as {\em fast reaction limit} in the following.

For reversal potentials $|E|\sim 0...100\,\rm mV$ and electrophoretic
charges of a few elementary units the parameter $\eta$ is in the
range $|\eta|\sim 0 \cdots 10$ under physiological conditions.
%
%
%
%
\section{Linear Stability Analysis}
\label{sec:lin}
We linearize  (\ref{eq:ma1}-\ref{eq:ma3}) around the homogeneous
and stationary solution 
(\ref{eq:stationary}) and apply the plane wave Ansatz~$(\varrho,\zeta,\Phi) = ( \varrho_k,\zeta_k,\Phi_k)\exp\{i{\mathbf
  k}\cdot{\mathbf x}+\omega t\} +c.c.$\\ As it was discussed above,
within the quasi-stationary approximation potential fluctuations
$\Phi_k$ are proportional to fluctuations of open channels
\begin{equation}
  \label{eq:phiqst}
  \Phi_k = -\eta\varrho_k/(1+k^2) \,.
\end{equation}
Inserting (\ref{eq:phiqst}) into the linearized (\ref{eq:ma1})
and (\ref{eq:ma2}) we get by abbreviating $\eta_k=\eta/(1+k^2)$ the eigenvalue problem
\begin{eqnarray}
  \label{eq:linearized}
\omega
\left(
\begin{array}{c}
    \varrho_k\\
    \zeta_k\\
\end{array}
\right)
 &=&
\left\lbrace
\left(
 \begin{array}{cc}
     -\gamma_-  & \gamma_-\\
     \gamma_+ & -\gamma_+ \\
   \end{array}
\right)\right. \nonumber \\
\nonumber \\
& &\left. 
-k^2
\left(
 \begin{array}{cc}
     1-\eta_k & 0\\
     -Dq\eta_k & -D\\
   \end{array}\
\right)
\right\rbrace
\left(
\begin{array}{c}
  \varrho_k\\
  \zeta_k
\end{array}
\right) \,.
\end{eqnarray}
The eigenvalues 
\begin{eqnarray}
  \label{eq:omegapm}
\omega_{\pm} &=& -[P(k^2)\pm\sqrt{Q(k^2)}]/2
\nonumber\\
\nonumber\\
P(k^2) &=& k^2(1+D-\eta_k)+\Gamma
\nonumber\\
\nonumber\\
Q(k^2) &=& \left[k^2(1-D-\eta_k)+\gamma_- - \gamma_+\right]^2
\nonumber \\
\nonumber\\
&& +\, 4Dq\gamma_-\eta_k + 4\,\gamma_-\gamma_+
\end{eqnarray}
determine linear stability. Let us note in passing that an extended
model with voltage dependent reaction rates will lead to the same set
of linearized equations.

The solution (\ref{eq:stationary}) becomes unstable against small
perturbations if ${\rm Re}\,\omega_+$ or ${\rm Re}\,\omega_-$ is
positive. Note that oscillatory unstable modes, which require ${\rm
  Im\,\omega_{\pm}} \neq 0$ and thus $Q<0$, are possible in the
parameter range $q\eta<0$.

The decisive influence of the parameters $\eta$ and $q$ on the
qualitative behavior of solutions is already apparent in the fully
non-linear system (\ref{eq:ma1}-\ref{eq:ma3}). The term
$\eta\varrho$ in (\ref{eq:ma3}) drives the curvature of the
potential $\Phi$, whereas $q$ may change the relative direction of
drift currents of the closed channels. For example, if open channels
carry negative charge, their drift is directed towards depolarized
regions of the membrane, where they accumulate. If a current through
open channels is directed into the thin layer of electrolyte, this
current will further depolarize the membrane and may cause an
instability. If closed channels have a positive effective
electrophoretic charge, the accumulated open channels will disperse
again after closing. This may lead to an oscillatory behavior. By
varying the relative signs of $q$ and $\eta$, different scenarios of
aggregation and dispersion of channels by lateral structures in $\Phi$
may arise, which are coupled back into the (\ref{eq:ma3}) and may
lead to stationary or oscillatory patterns.

In the following we will hence discuss the four parameter regions
${\mathcal{R}_{++}}\widehat{=}\lbrace q>0,\eta>0\rbrace$,
${\mathcal{R_{--}}}\widehat{=}\lbrace q<0,\eta<0\rbrace$,
${\mathcal{R_{+-}}}\widehat{=}\lbrace q>0,\eta<0\rbrace$ and
${\mathcal{R_{-+}}}\widehat{=}\lbrace q<0,\eta>0\rbrace$ separately.
The quantity $\eta$ is treated as the primary control parameter.

\subsection{Instabilities for $q\eta>0$}
\label{sec:statpat}

In these regions of parameters, $Q(k^2)$ is always positive, the rates
$\omega_{\pm}(k^2)$ are real and thus plane wave solutions will either
exponentially grow or decay in time. Instabilities signal the onset of
stationary patterns.  As $\omega_+>\omega_-$, it is the rate
$\omega_+$, which will drive the solution (\ref{eq:stationary})
unstable, if control parameters reach the stability boundary. Note
that in the long wavelength limit $k \to 0$, $\omega_+$ vanishes for
all values of $\eta, q, D, \gamma_{\pm} $ like
$\omega_+=O(k^2)$. The corresponding diffusion mode reflects
conservation of the total number of channels.
  
From 
$\omega_+ \omega_- =0$ we get the neutrality condition
\begin{equation}
  \label{eq:tve0}
\eta_0(k) = (1+k^2)\frac{D(k^2 + \gamma_{-}) + \gamma_+}{D(k^2 + q\gamma_-) + \gamma_+}\,.
\end{equation}
The minimum of $\eta_0(k)$ corresponds to the first onset of
instabilities and the corresponding wavenumber $k_c$ is the first
unstable mode (critical mode).  A minimum of the neutral
curve appears at $k_c\neq 0$ with 
\begin{equation}
  \label{eq:kcq}
k_c^2 = -q\gamma_- - \frac{\gamma_+}{D} +\sqrt{\gamma_-(q-1)(q\gamma_- + \gamma_+/D -1)} 
\end{equation}
if the {\em rhs} of (\ref{eq:kcq}) is a positive, real number. By
varying $q, D$ or the rates $\gamma_{\pm}$, the instability may switch
between soft-mode ($k_c=0$) and hard-mode ($k_c\neq 0$) type.

Several statements are noticeable: (a) For $D=0$, (immobile closed
channels), as well as for $q=1$, the neutrality condition,
$\eta_0=1+k^2$, is independent of the reaction rates and implies a
soft mode instability at $\eta_c=1$. A hard mode instability requires
both mobile channels and different charges of closed and open states.
(b) Hard mode instabilities are only possible for $q<1$, i.e.
$q_c<q_o$. (c) For $D\to\infty$ and $q=0$ we recover the result of
\cite{fromherz1995a}. (d) In the fast reaction limit
($\gamma_->>1, \gamma_+>>1, D\neq 0$), only soft mode instabilities at
$\eta_c=(D\gamma_-/\gamma_+ + 1)/(Dq\gamma_-/\gamma_+ +1)$ remain.
Note that the critical control parameter stays small ($O(1)$) in the
fast reaction limit, indicating that the instability may indeed be
reached under physiological conditions.

For reaction rates of order 1, typical curves of $\eta_0(k)$ are shown
in Fig.\ref{fig:nkurve++}.  Whereas in ${\mathcal R_{++}}$, both
soft and hard mode instabilities may occur, a neutral curve in
${\mathcal R_{--}}$ does only appear for $D|q|>\gamma_+/\gamma_-$, in
which case $\eta_0(k)$ is maximal at $k=0$ and exhibits poles at
$k_p^2=-q\gamma_--\gamma_+/D$. Thus, for electrophoretic charges $q_o$
and $q_c$ of different signs, merely soft-mode instabilities at
$\eta_c < 0$ are possible as it is demonstrated in
Fig.\ref{fig:nkurve++}.b). Note that only a finite band of unstable
modes, limited to the interval $(-k_p,k_p)$, will appear for all
$\eta<0$. This unstable band grows like $k_p\sim\Gamma^{1/2}$ in the
fast reaction limit.
Fig.\ref{fig:hardorsoft} shows the $k_c^2=0$ lines in the
$D$-$\Gamma)$-plane for different values of $q$. For parameters
$(D,\Gamma)$ below each curve $k_c^2\neq 0$.

Setting $\varrho_{k_c}=1$ and solving the linearized system for the
eigenvector $(\varrho_{k_c}, \zeta_{k_c}, \Phi_{k_c})$ of the critical
mode, we obtain
$\zeta_{k_c}=(qk_c^2+q\gamma_-+\gamma_+/D)/(k_c^2+q\gamma_-+\gamma_+/D)$
and
$\Phi_{k_c}=-(k_c^2+q\gamma_-+\gamma_+/D)/(k_c^2+q\gamma_-+\gamma_+/D)$.
Taking a closer look at the signs we find ${\rm sgn}\, \varrho_{k_c}=
{\rm sgn}\, \zeta_{k_c}$ and ${\rm sgn}\, \varrho_{k_c}= -{\rm sgn}\,
\Phi_{k_c}$ if $q>0$. Hence for $q>0$, spatial variations of open and
closed channels are in phase but have a phase lag of $\pi$ with
respect to the potential in the critical mode. But if $q<0$ there is
no phase lag, as then ${\rm sgn}\, \omega_{k_c}= {\rm sgn}\,
\zeta_{k_c} = {\rm sgn}\, \Phi_{k_c}$. In this case the critical
wavenumber vanishes leading to $\zeta_0=1$ and $\Phi_0=-\eta_c$ so that all spatial
periods are in phase as expected from the qualitative discussion on
the mechanism of pattern formation given in (\ref{sec:model}) after
(\ref{eq:ma3}).


\subsection{Instabilities for $q\eta<0$}
\label{sec:oscpat}
The simplest behavior of the system appears in the regime ${\mathcal
  R_{+-}}$.  For ${\rm Im} \omega_{\pm}\neq 0$, it is obvious from
(\ref{eq:omegapm}) that ${\rm Re} \omega_{\pm}=-(1/2)P(k^2)<0$. It
is easy to show that linear stability also holds for ${\rm Im}
\omega_{\pm}=0$ and thus the homogeneous, stationary solution
(\ref{eq:stationary}) is linearly stable in this regime.

It is only in ${\mathcal R_{-+}}$, that oscillatory instabilities may
arise, because $Q(k^2)$ and $P(k^2)$ can simultaneously take on
negative values in this regime. If $Q<0$, $\omega_+=\omega_-^*$, and
the growth rate of the perturbation is ${\rm Re} \omega_{\pm} =
-P(k^2)/2$.
From the condition $P(k^2,\eta) =0$ we obtain the neutrality condition
$\eta_P(k)$ for oscillatory instabilities
\begin{equation}
  \label{eq:nko}
  \eta_P(k) = (1+D)(1+k^2)\left( 1 + \frac{\Gamma}{(1+D)k^2}\right)\,.
\end{equation}
It attains its minimum at the critical wavenumber
\begin{equation}
  \label{eq:kpc}
 k_{P_{c}}=\,\, ^4\sqrt{\frac{\Gamma}{1+D}} \,.
\end{equation}
The critical value of the control parameter is 
\begin{equation}
  \label{eq:epm}
  \eta_{P_{c}}= (1+D)\left( 1 + \sqrt{\frac{\Gamma}{1+D}} \right)^2 \,. 
\end{equation}
Note that $P(k^2)$ does not depend on the ratio $q$ of the
electrophoretic charges and thus the neutral curve only depends on
reaction rates and diffusion coefficients.  The ratio $q$ will,
however, show up in the oscillation frequency
$\Omega_c=(1/2)\sqrt{|Q(k_c^2)|}$ at the critical point.

The region ${\mathcal R_{-+}}$ also contains parts with $Q(k^2)>0$,
where stationary unstable patterns may arise if the neutral curve
determined from (\ref{eq:tve0}) is crossed. Hence the true critical
point is located at
\begin{equation}
  \label{eq:etacinrpm}
  \eta_c=\min_k\{ \eta_0(k), \eta_P(k)\} \,.
\end{equation}
The oscillatory instabilities will not be attainable in the fast
reaction limit under physiological conditions because $\eta_{P_c}\sim
\Gamma$ for large $\Gamma$.

Fig.\ref{fig:nkurve-+} shows neutral curves in ${\mathcal R_{-+}}$
together with the region $Q(k^2)<0$ for two cases, which differ by the
value of $D$. In Fig.\ref{fig:nkurve-+}.a, the unstable pattern is
oscillating, in Fig.\ref{fig:nkurve-+}.b, it is stationary.
Fig.\ref{fig:criticalomega} shows the dependence of the critical
frequency $\Omega_c$ on $D$ and $\Gamma$ for different $q$.
Note, that by decreasing $D$, the oscillations will set in with a
finite frequency at the transition, whereas by increasing $D$ the
critical frequency $\Omega_c$ starts out from $\Omega_c=0$ at the transition.

\section{Effects in the weakly nonlinear regime }
\label{sec:wnl}
We now turn to a discussion of the behavior of the system beyond
linear stability analysis. This discussion is restricted to
stationary patterns. A more detailed analysis of oscillatory patterns
will be given elsewhere.

In the vicinity of the critical control parameter $\eta_c$ and the
critical wavenumber $k_c$ the linear dispersion of the slow mode
$\omega_{+}(\eta, k^2)$ may be expanded to obtain the linear part of a
non-linear evolution equation for the slow mode near the bifurcation
point.  From (\ref{eq:omegapm}) one gets
\begin{eqnarray}
  \label{eq:expand.omega.soft} 
\omega_{+}(\eta, k^2) & = & k^{z_a-2}
\tau_a^{-1}\Bigg\{\frac{\eta-\eta_c}{\eta_c}\ +\ \xi_a^2\,(k-k_c)^2 
  \nonumber \\
& &
 + O[(k-k_c)^4]\Bigg\}\,,
\end{eqnarray}
where $z_a=4$ for soft-mode instabilities ($a=s$) and $z_a=2$ for
hard-mode instabilities ($a=h$). The expressions for time scales
$\tau_a$ and length scales $\xi_a$ can be found in Appendix A for
soft-mode and in Appendix B for hard-mode instabilities.

\subsection{Soft-Mode Instabilities}
\label{subsec:wnl.soft}
To proceed beyond the linearized dynamics near a soft-mode
instability, it is convenient to introduce the variable
$u=(\gamma_+\varrho+\gamma_-\zeta)/\Gamma$, which corresponds to
fluctuations of the total channel density, $u=
(n_o+n_c-\bar{n})/\bar{n}$, and which contains the critical slow mode
as $\eta \to \eta_c$. Furthermore we introduce
$v=(\varrho-\zeta)/\Gamma$, which remains fast relative to $u$ near
the bifurcation. We get an equation for the slowly varying part of
$u(x,t)$ from a gradient expansion, which is outlined in Appendix A.
 
If we take into account all terms of $O(\nabla^2)$ and include the
$O(\nabla^4)$ terms, which arise from the linear dispersion
(\ref{eq:expand.omega.soft}), the resulting equation for $u$ takes
on the form of a dynamical Cahn-Hilliard ( or 'model B') equation
\begin{equation}
  \label{eq:amplitude.soft}
  \partial_tu = \tau_s^{-1}\nabla^2\left\{\frac{\delta F(u)}{\delta u} \right\} + O(\nabla^4u^2)\,,
\end{equation}
with an effective potential $F(u)=\int[f(u)+(\xi_s^2/2)(\nabla
u)^2]d^2{\bf r}$.\\ The local part $f(u)$ is given by
\begin{eqnarray}
  \label{eq:softpotential}
  f(u)&=&\frac{\eta_c
    u^2}{2}+\frac{\eta}{\alpha^2}\Bigg\lbrack(2-\alpha)u \nonumber \\
\\
& & +\left(1- \frac{2}{\alpha}
      -u\right)\ln(1+\alpha u)  \Bigg\rbrack\,.\nonumber
\end{eqnarray}
Equation (\ref{eq:amplitude.soft}) has to be solved under the constraint of
conservation of channel number, $\int\, u\, d^2{\bf r}=0$, which may
be taken into account by an appropriate Lagrange multiplier term
$-\Lambda u$ in $f(u)$.  Note that the dynamics only depends on the
ratio $\eta_c/\eta$ and on the parameter $\alpha$ and thus exhibits a
particularly simple universality for all permissible values of
$D,q,\gamma_{\pm}$.

Fig.\ref{fig:potential} shows the effective potentials for
$\alpha=0.6$ close to the transition.  $\eta_c$ becomes the spinodal.
The discussion of the dynamical behavior, including typical coarsening
in the case of quenches can be found in standard textbooks, see for instance
\cite{chaikin1995}.

\subsection{Hard-Mode Instabilities}
\label{subsec:wnl.hard} 
In order to obtain an analytic description of the weakly nonlinear
regime for hard-mode instabilities, we use a standard multi-scale
perturbative approach, which results in an amplitude equation
\cite{cross1993}. The approach starts from the Ansatz
\begin{equation}
  \label{eq:lin_u}
{\bf u}_c
\left[\chi(X,Y,T)e^{i k_c x} +
  \chi^{*}(X,Y,T)e^{-i k_c x}\right] \quad 
\end{equation}
for the fluctuations ${\bf u}=(\varrho,\zeta,\Phi)$. The amplitude
$\chi$ depends upon the scaled variables $X=a x$, $Y=a^{1/2} y$ and
$T=a^2t$. The scale $a$ becomes small, if the control parameter $\eta$
approaches its critical value $\eta_c$ from above,
$a^2=(\eta-\eta_c)/\eta_c.$ For convenience we sketch the procedure in
Appendix B.
For one-dimensional, stationary patterns the perturbation theory gives
for the amplitude $\chi$ in the supercritical vicinity of $\eta_c$ the
equation
\begin{equation}
  \label{eq:amplitude}
  \tau_h\partial_t\chi = \frac{\eta-\eta_c}{\eta_c}\chi +\xi_h^2\left\lbrack\partial_x -\frac{i}{2k_c}\partial_y^2
  \right\rbrack^2\chi -J|\chi|^2\chi \,.
\end{equation}
The lengthy expression, which gives $J$ as a function of the model
parameters, is given in the Appendix B. Fig.\ref{fig:J} displays the
sign of $J$ in the $D$-$\Gamma$-plane for different values of
$\alpha$. 
Fig.\ref{fig:J} shows the distribution of the sign of $J$ within the
$D$-$\Gamma$-plane for different values of $q$ and $\alpha$. Positive
$J$ corresponds to a continuous transition (forward bifurcation)
negative $J$ to a discontinuous one (backward bifurcation). 

Note that for $\alpha\le0.5$, Fig.\ref{fig:J}.a and Fig \ref{fig:J}.b,
there is inside of the domain of $k_c^2>0$ only one region of positive
$J$, i.e. continuous transitions, which is generically bounded by the
domain of negative $J$, i.e. discontinuous transitions. In the other
regime $\alpha>0.5$, Fig.\ref{fig:J}.c and Fig \ref{fig:J}.d, the
system shows a reentrant behavior, as then an additional domain with
$J>0$ appears below the domain of $J<0$. This means, coming from large
values of $D$ one leaves the area of continuous transitions, crosses
the domain of discontinuous transitions and again enters a domain with
$J>0$, before finally the $k_c^2=0$-line is crossed.  Parallel to the
$\Gamma$ axis the generic behavior is similar.

\section{Conclusions}
\label{sec:con}
We have studied a simple model of spontaneous pattern formation of
mobile ions in a fluid membrane. The channel proteins may switch
between an open and a closed state according to a simple two state
reaction kinetics with constant rates. The effective electrophoretic
charge and the diffusion constant of the channel proteins are assumed
to be state dependent. The reversal potential $E$ of ions, which may
pass the open channels constitutes a non-equilibrium driving
force. The characterization of a particular model is completed by the
additional parameters $D=D_c/D_o$ and $q=q_c/q_o,$ the reaction rates
$\gamma_{\pm}$ and the fraction $\alpha$ of conductance due to the
average number of open channels to the average total transmembrane
conductivity. Varying these parameters leads to a number of distinct
scenarios for first bifurcations from the stationary and homogeneous
state.  Depending on the sign of the control parameter
$\eta=\alpha(\alpha-1)\beta q_oE$ and the ratio $q=q_c/q_o$ of
electrophoretic charges, four qualitatively different regions of
parameters have been identified from linear stability analysis: (a) for
$q>0$ and $\eta>0$ (${\cal R}_{++}$) soft- or hard mode instabilities
leading to stationary patterns will appear, (b) for $q<0$ and $\eta<0$
(${\cal R}_{--}$) only bifurcations with soft mode instabilities can
occur, (c) for $q<0$ and $\eta>0$ (${\cal R}_{-+}$) hard-mode
instabilities with or without temporal oscillations may be found and
(d) for $q>0$ and $\eta<0$ (${\cal R}_{+-}$) the homogeneous,
stationary state remains linearly stable.  In the fast reaction limit,
where the time scale $\Gamma^{-1}$ becomes much shorter than other
time scales (except the RC-relaxation time of transmembrane potential
fluctuations), only soft-mode instabilities remain.  By varying the
ratio $D=D_c/D_o$ of diffusion constants or $q=q_c/q_o$ of effective
electrophoretic charges within physiologically plausible bounds, it is
possible to switch between soft- and hard-mode instability in ${\cal
  R}_{++}$ and between hard-mode and oscillatory instability in ${\cal
  R}_{-+}$.

For soft-mode instabilities, we have derived an equation of motion of
slow modes near the transition, which is of the form of a dynamical
Cahn-Hilliard equation. The transition will generically be
discontinuous (backward bifurcation) and the non-linear evolution will
be characterized by regions of nucleation and of spinodal
decomposition. Quenches into the supercritical regime are predicted to
show coarsening behavior with a coarsening length scale growing as
$t^{1/3}$.  For hard-mode instabilities, an amplitude equation is
obtained as in \cite{fromherz1995a} and parameter regions of
forward and of backward bifurcations may be distinguished by the sign of
the non-linear coupling parameter.

The presented model includes and extends several previously studied
models and may be considered as a minimal model for charged ion
channels with internal state kinetics in a fluid membrane.

\begin{appendix}
\section{Gradient Expansion near Soft-Mode Instabilities}
We transform  (\ref{eq:ma1}-\ref{eq:ma3}) to the new variables
\begin{eqnarray}
u & = &(\gamma_+\varrho+\gamma_-\zeta)/\Gamma
\nonumber\\
\nonumber\\
v & = & (\varrho - \zeta)/\Gamma
\end{eqnarray} 
and get
\begin{eqnarray}
\label{eq:u}
\Gamma\partial_t u & = & \quad (\gamma_+ + D\gamma_-)\nabla^2u+\gamma_+\gamma_-(1-D)\nabla^2v +
\nonumber\\
\nonumber\\
&&+\, (\gamma_+ + r\gamma_-)\nabla^2\Phi+ (\gamma_+ +
r\gamma_-)\nabla(u\nabla\Phi)\nonumber\\ 
\nonumber\\
&& +\, \gamma_+\gamma_-\nabla(v\nabla\Phi)
\end{eqnarray}
as equation of motion for $u$. Note that all terms on the
{\em rhs} of the equation for $u$ are at least $O(\nabla^2)$ as a direct consequence of channel number conservation.
For $v$ we get
\begin{eqnarray}
\label{eq:v}
\Gamma\partial_t v & = & -\Gamma^2 v+ (\gamma_- + D\gamma_+)\nabla^2v+(1-D)\nabla^2u +
\nonumber\\
\nonumber\\
&&+ (1-r)\nabla^2\Phi + (1 - r)\nabla(u\nabla\Phi) \nonumber \\
\nonumber\\
&&+(\gamma_- + r\gamma_+)\nabla(v\nabla\Phi)
\end{eqnarray}
Note that $v$ decays with a rate $O(\nabla^0)$ due to the $-\Gamma^2v$ term.
Thus $v$ is fast near a soft-mode transition and
may be adiabatically eliminated. It does not enter the leading order of the gradient expansion, which is obtained by
replacing $\Phi$ in (\ref{eq:u}) from the leading $(O(\nabla^0))$ order of 
\begin{eqnarray}
\label{eq:Phi}
0 & = & (1-\nabla^2)\Phi+\eta(u+\gamma_-v)+\alpha(u+\gamma_-v)\Phi\,.
\end{eqnarray}
This leads to
\begin{equation}
\label{eq:done}
\partial_t u = \tau_s^{-1}\nabla^2\left\{ \eta_c u +(1+u)\phi(u) -\int\phi(u)\,du)\right\}
\end{equation}
with $\tau_s^{-1} = (\gamma_+ + D\gamma_-)/\Gamma$ and $\phi(u)=-\eta
u/(1+\alpha u)$.

The linear $O(\nabla^4)$ term is obtained from the expansion of
$\omega_+(\eta,k^2) =
\tau_s^{-1}[k^2(\eta-\eta_c)/\eta_c+\xi_s^2k^2+O(k^4)]$. For $\xi_s^2$
we obtain
\begin{eqnarray}
\frac{\xi_s^2}{\tau_s} & = & -\frac{\eta}{2} + \frac{1}{4\Gamma}\left\lbrack(1-D-\eta)^2+2\eta(\gamma_- - \gamma_+) - 4\eta\gamma_-Dq\right\rbrack
\nonumber\\
&& - \frac{1}{4\Gamma^3}\left\lbrack(\gamma_- - \gamma_+)(1-D-\eta) + 2 \eta\gamma_-Dq\right\rbrack^2
\end{eqnarray}
Adding a term $-\tau_s^{-1}\xi_s^2\nabla^4u$ on the {\em rhs} of
(\ref{eq:done}) we get (\ref{eq:amplitude.soft}) and
(\ref{eq:softpotential}).

%
\section{Amplitude Equation near Hard-Mode Instabilities}
The system of nonlinear partial differential equations
(\ref{eq:ma1}-\ref{eq:ma3}) is written as $ {\cal L}{\bf u} = {\cal
  N}({\bf u},{\bf u})$ In the vicinity of $\eta_c, k_c$ ${\bf u} =
(\varrho, \zeta, \Phi)$ contains both fast scale variations (with
$\nabla u=O(k_c)$) and slow scale variations (with $\nabla u \to 0$
for $\eta \to \eta_c$). The slowly varying part may be obtained from a
perturbation expansion in $\epsilon = \sqrt{|(\eta -
  \eta_c)/\eta_c|}$, which explicitly separates slow scale variations
from fast scales by introducing appropriately scaled length and time
variables $X=\epsilon x$ $T=\epsilon^2 t$.  For one-dimensional
patterns, the direction perpendicular to the pattern wave vector
should be scaled as $Y=\epsilon^{1/2}y$ \cite{newell1969}. As
functions now may depend upon fast and slow variables, we have to
replace $\partial_t \to \partial_t+\epsilon^2\partial_T$ and
$\partial_x\to\partial_x+\epsilon\partial_X$ etc. In this way, the
linear part ${\cal L}$ becomes ${\cal L} = {\cal L}_0 + \epsilon{\cal
  L}_1 + \epsilon^2{\cal L}_2$. Inserting the expansion ${\bf
  u}=\epsilon {\bf u}_1 + \epsilon^2{\bf u}_2 +\epsilon^3{\bf
  u}_3+O(\epsilon^4)$ (with ${\bf
  u}_1(X,Y,T,x)=(\chi(X,Y,T)\exp(ik_cx)\, +\, cc$) and sorting with
respect to powers of $\epsilon$ one gets a hierarchy of equations,
which allows to determine ${\bf u}_n$ from the ${\bf u}_k$ with $k<n$.
It starts with ${\cal L}_0{\bf u}_1=0$, which implies that ${\bf u}_1$
is an eigenstate to ${\cal L}_0$ with eigenvalue zero. As $\partial_T$
carries an explicit factor of $\epsilon^2$, the equation, which
contains the slow time derivative of ${\bf u}_1$ appears in third
order in $\epsilon$ and has the form ${\cal L}_2{\bf u}_1 + {\cal
  L}_1{\bf u}_2 + {\cal L}_0{\bf u}_3 = {\cal N}_3({\bf u}_1,{\bf
  u}_2)$. Inserting ${\bf u}_2$ as obtained from the perturbation
expansion in terms of ${\bf u}_1$, one gets the amplitude equation by
taking the scalar product of the third order equation with the left
eigenstate of ${\cal L}_0$ to eigenvalue zero, ${\mathbf
  u}_1^{\dagger}$. This solvability condition gives
\begin{eqnarray}
\label{eq:ampl}
  \int{\bf u}_1^{\dagger} 
{\cal L}_2{\bf u}_1d^2{\bf r} +  \int {\bf u}_1^{\dagger} 
{\cal L}_1{\bf u}_2 d^2{\bf r}
  = \int {\bf u}_1^{\dagger} {\cal N}_3 d^2{\bf r}
\end{eqnarray}
The {\em lhs} of (\ref{eq:ampl}) is linear in $\chi$ and can be
obtained directly from expanding $\omega_+(\eta, k^2)$ around $\eta_c,
k_c$, $\omega_+(\eta, k^2)=\tau_h^{-1}\{\epsilon^2+\xi_h^2(k-k_c)^2 +
O[(k-k_c)^4]\}$. Inserting this expansion on the {\em lhs} of
(\ref{eq:ampl}) and neglecting the $y$ dependence for
simplicity gives
\begin{equation}
\label{eq:ampldone}
\left(\tau_h\partial_t-
  \frac{\eta-\eta_c}{\eta_c}-\xi_h^2\partial_x^2\right)\chi =  \int
{\bf u}_1^{\dagger} {\cal N}_3d^2{\bf r} 
\end{equation} 
For the time constant $\tau_h$ we get
\begin{equation}
\label{tau0}
\tau_h = \frac{\Gamma + (1+D)k_c^2 -\eta_c k_c^2/( 1+k_c^2)}
{Dk_c^2(k_c^2 +\gamma_- + \gamma_+/D)}
\end{equation} 
and the square $\xi_h^2$ of the correlation length is 
\begin{equation}
\label{xi0}
\xi_h^2 = \frac{4 k_c^2}{(1+k_c^2)\left(k_c^2 +\gamma_- +\frac{\gamma_+}{D}\right)} \,.
\end{equation}
Evaluating the non-linear term on the {\em rhs} of
(\ref{eq:ampldone}) gives (\ref{eq:amplitude}) with the
following expression for J
\begin{eqnarray}
J &=& 
\frac{-\eta_c}{(k_c^2 + \gamma_- +\frac{\gamma_+}{D} )(1 + k_c^2)^2}
\bigg\lbrace  
\alpha \left( q\gamma_- +\frac{\gamma_+}{D} + k_c^2
\right)\cdot\nonumber \\
\nonumber\\
& &\cdot (p_1 +\Phi_{k_c}n_1 +2\alpha)%
\nonumber\\
\nonumber \\  
&& -(1 + k_c^2)\Big\lbrack 2p_1\left(\frac{\gamma_+}{D} + k_c^2 + \zeta_{k_c}q\gamma_-\right) \nonumber \\ 
\nonumber \\ 
&&
-\Phi_{k_c}\left\lbrace n_1(\gamma_+ + k_c^2)  -q\gamma_- (1 + k_c^2)z_1\right\rbrace \Big\rbrack
\bigg\rbrace \,.\nonumber 
\end{eqnarray}
The expressions $n_1$, $z_1$ and $p_1$ stem from ${\cal O}(\epsilon^2)$
corrections and have the form
\begin{eqnarray}
  \label{eq:n1}
  n_1 &=& \frac{1}{18\,k_c^4} \bigg\lbrace  (1+4k_c^2)(\gamma_+/D 
  +4k_c^2 +\zeta_{k_c}q\gamma_-) \nonumber \\& &\qquad\qquad -2\alpha (\gamma_+ + q\gamma_- +4k_c^2)
  \bigg\rbrace\,,\\ 
\nonumber \\
 \label{eq:z1}  
 z_1 &=& \frac{ n_1(\beta+2k_c^2)(1 +4k_c^2) +4k_c^2(\alpha -n_1\eta_c)}{\beta(1+4k_c^2)(1
  +k_c^2)}\,,\\
\nonumber \\
  \label{eq:p1}
  p_1 &=& \frac{\left( 1+k_c^2\right) En_1 +\alpha}{1 +4k_c^2} \,.
\end{eqnarray}
\end{appendix}
%
%
%
%
 

%

\newcommand{\captionFig}{Different biological models. Left: simplified synapse. Right: Model of an axon.}

\newcommand{\captionNeutraleKurvenabc}{Neutral curves in regions
  ${\cal R}_{++}$ (a), and ${\cal R}_{--}$ (b) for different values of
  q. For the upper, middle and lower curve $q=0.1,\ 0.2$ and $0.6$ in
  (a), $q=-3.0,\ -2.0$ and $-1.0$ in (b) and $q=-0.3, \ -0.5, \ -1.0$.
  Other parameters are $\gamma_+=0.01,\ \gamma_-= 0.5$ and $D=0.5$ }

\newcommand{\captionNeutraleKurvenmitImab}{Neutral curves and domain
  of complex eigenvalues (shaded) in ${\cal R}_{-+}$ for small
  $D=0.09$ (a), and $D=5.0$ (b). Note that there is a discontinuous
  change in the nature of the transition with increasing $D$; from
  hard-mode, stationary to hard-mode, oscillatory type. Other
  parameter values are $q=-2.0$,\ $\gamma_+=0.01,\ \gamma_-=0.5$.}

\newcommand{\captionGrenzeHardModeSoftMode}{Boundaries between
  hard-mode and soft-mode instabilities in $D-\gamma$-plane in ${\cal
    R}_{++}$ for different values of $q$. In regions below each curve
  $k_c^2\neq 0$.  From lower to upper curves, values of $q$ are $0.0,\ 
  0.1,\ 0.25$ and $0.5$ from lower to upper curves.}

\newcommand{\captionCriticalomega}{The critical frequency $\Omega_c$:
  a) As function of $D$ for $\Gamma = 4.0$, 4.5, 5.0 and b) as a
  function of $\Gamma$ for $D=2.75$. The remaining parameters are
  $\gamma_+-\gamma_- = 1$ and $q= -2$.}

\newcommand{\captionSoftmodePotential}{Local part of the effective
  potential $f(u)$, which controls the Cahn-Hilliard dynamics near
  soft-mode instabilities at $\alpha=0.6$. The uppermost curve
  displays the situation at the transition ($\eta_c/\eta=1.02525$),
  whereas the middle curve corresponds to the spinodal point
  $\eta_c/\eta=1$. The lower curve is in regime at $\eta/\eta_c=0.97$.
  For better visibility, the $f(u)$ at the transition point has been
  enlarged by a factor of 10. If $\alpha<0.5$ the global minimum of
  the potential is attained for positive $u$ and and the graphs look
  like mirrored at the $u=0$-axis.}

\newcommand{\captionJDG}{$J=0$-lines within the $k_c^2\ge 0$-domain
  for $g=0.1$, $\gamma=2.0$ and various $\alpha$: (a) 0.25, (b) 0.5,
  (c) 0.6 and (d) 0.75. The value of $\alpha$ crucially determines the
  number of domains with continuous transitions (shaded). If
  $\alpha>0.5$ the system clearly displays reentrant behavior.}

\begin{figure}
\mbox{}
\hfill\parbox[b]{0.45\linewidth}
{\epsfig{file=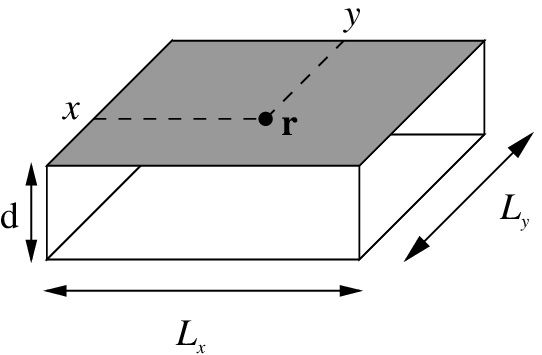, width=\linewidth}}
\parbox[b]{0.45\linewidth}
{\epsfig{file=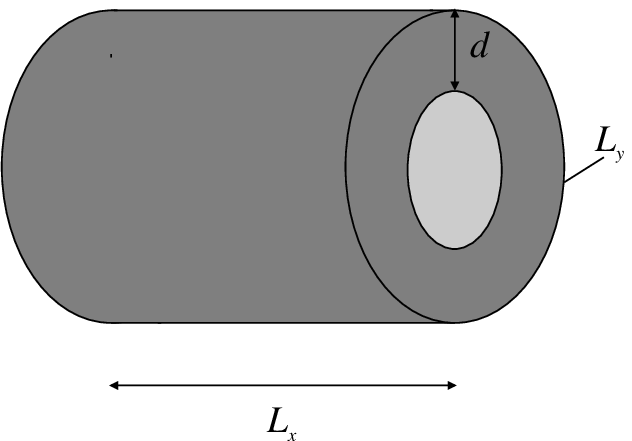, width=\linewidth}}
\hfill\mbox{}
\caption{\captionFig}
   \label{fig:fig1}
\end{figure}


\begin{figure}\begin{center}
\mbox{}
\hfill\parbox[b]{0.45\linewidth}
{\epsfig{file=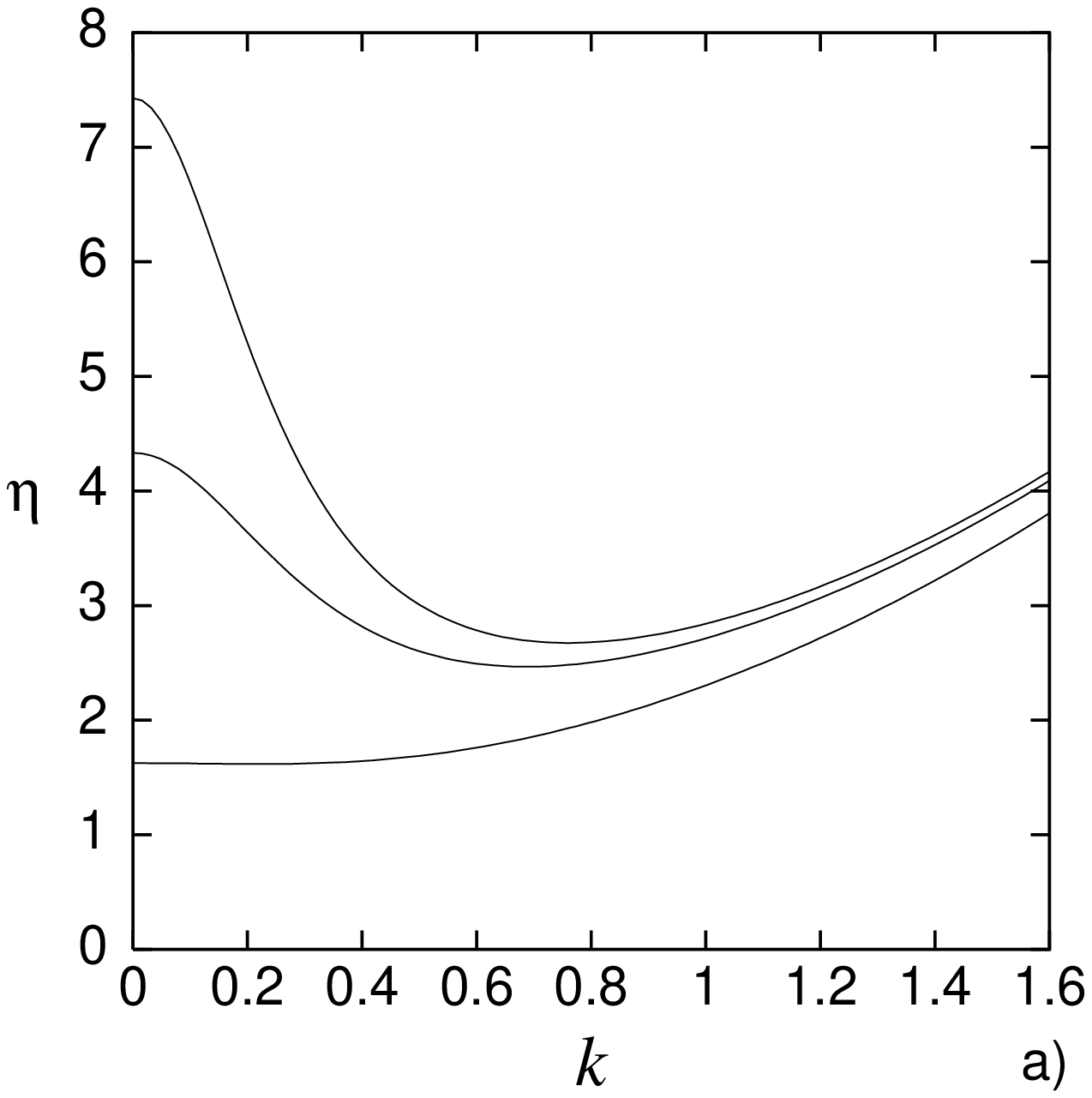, width=\linewidth}}
\parbox[b]{0.45\linewidth}
{\epsfig{file=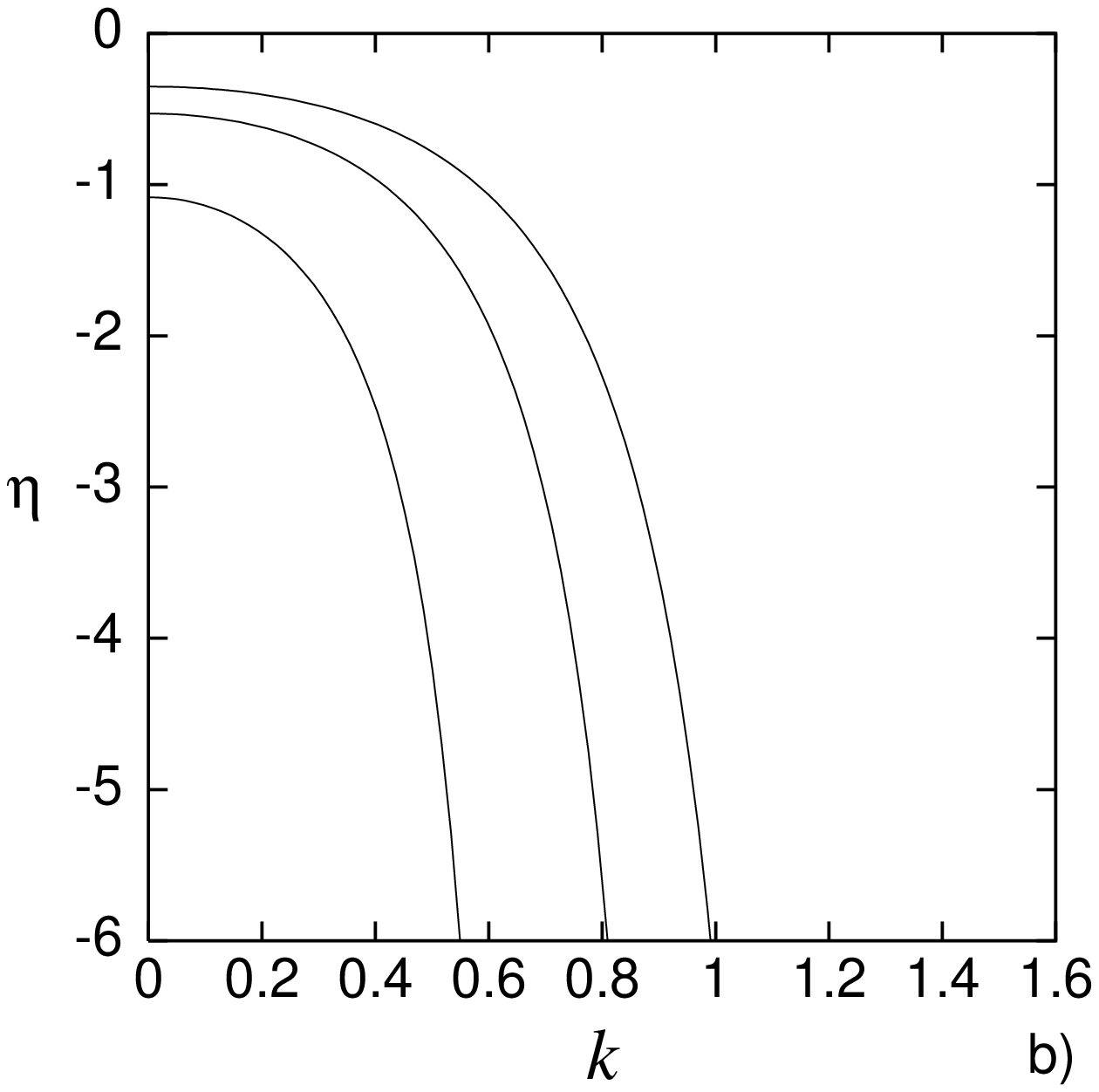, width=\linewidth}}
\hfill\mbox{}\\
 \end{center}
 \begin{center}
  \mbox{}
  \hfill
\parbox[b]{0.45\linewidth}{\epsfig{file=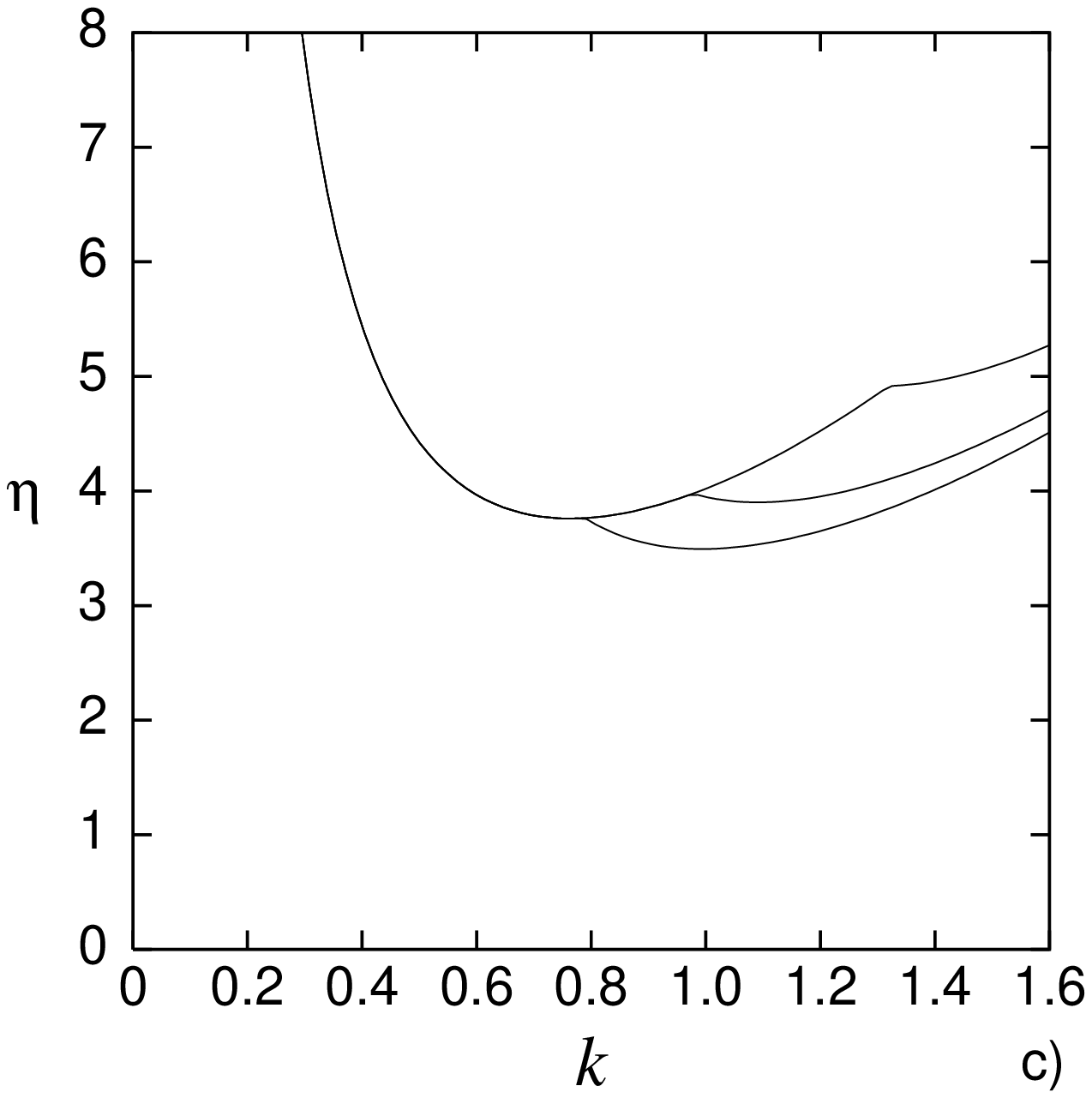, width=\linewidth}}
  \hfill\mbox{} 
 \end{center} 
\caption{\captionNeutraleKurvenabc}
\label{fig:nkurve++}
\end{figure}


\begin{figure}
\epsfig{file=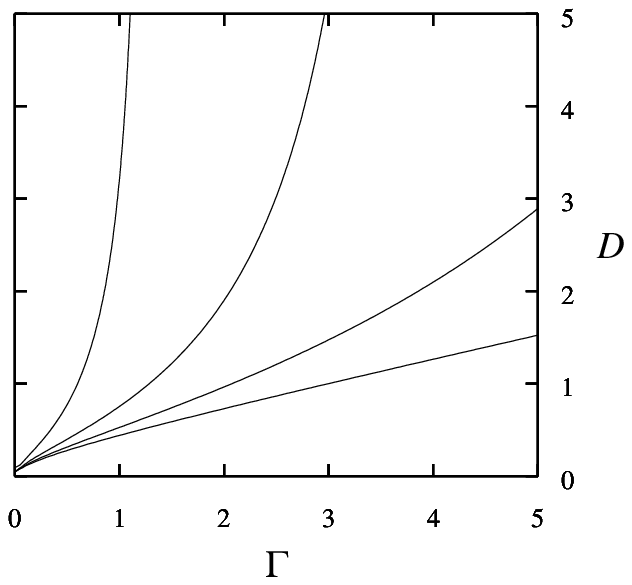, width=\linewidth}
\caption{\captionGrenzeHardModeSoftMode}
\label{fig:hardorsoft}
\end{figure}

\begin{figure}[hbt]
\mbox{}
\hfill\parbox[b]{\linewidth}
{
\epsfig{file=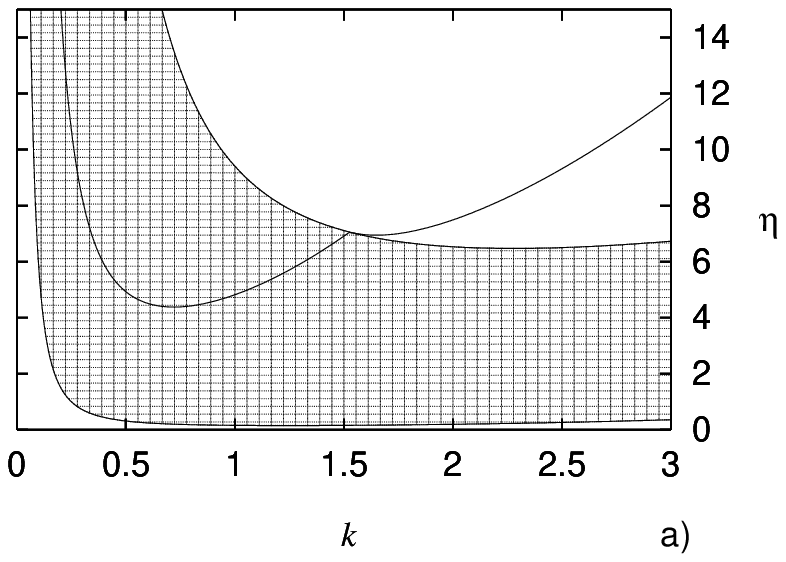, width=1.2\linewidth}
}
\parbox[b]{\linewidth}
{
\epsfig{file=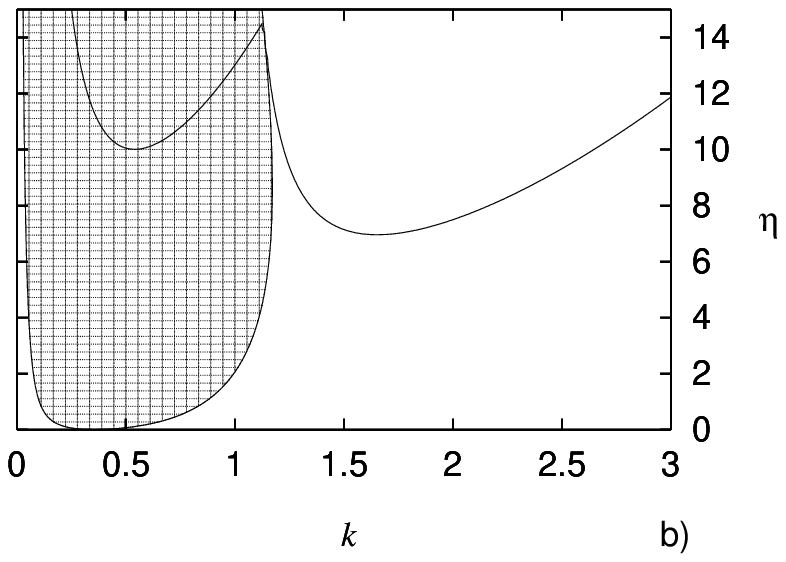, width=1.2\linewidth}
}
\hfill\mbox{}
\caption{\captionNeutraleKurvenmitImab}
\label{fig:nkurve-+}
\end{figure}

\begin{figure}[htbp]
\begin{center} 
\mbox{}
 \hfill\parbox[b]{\linewidth}
 {\epsfig{file=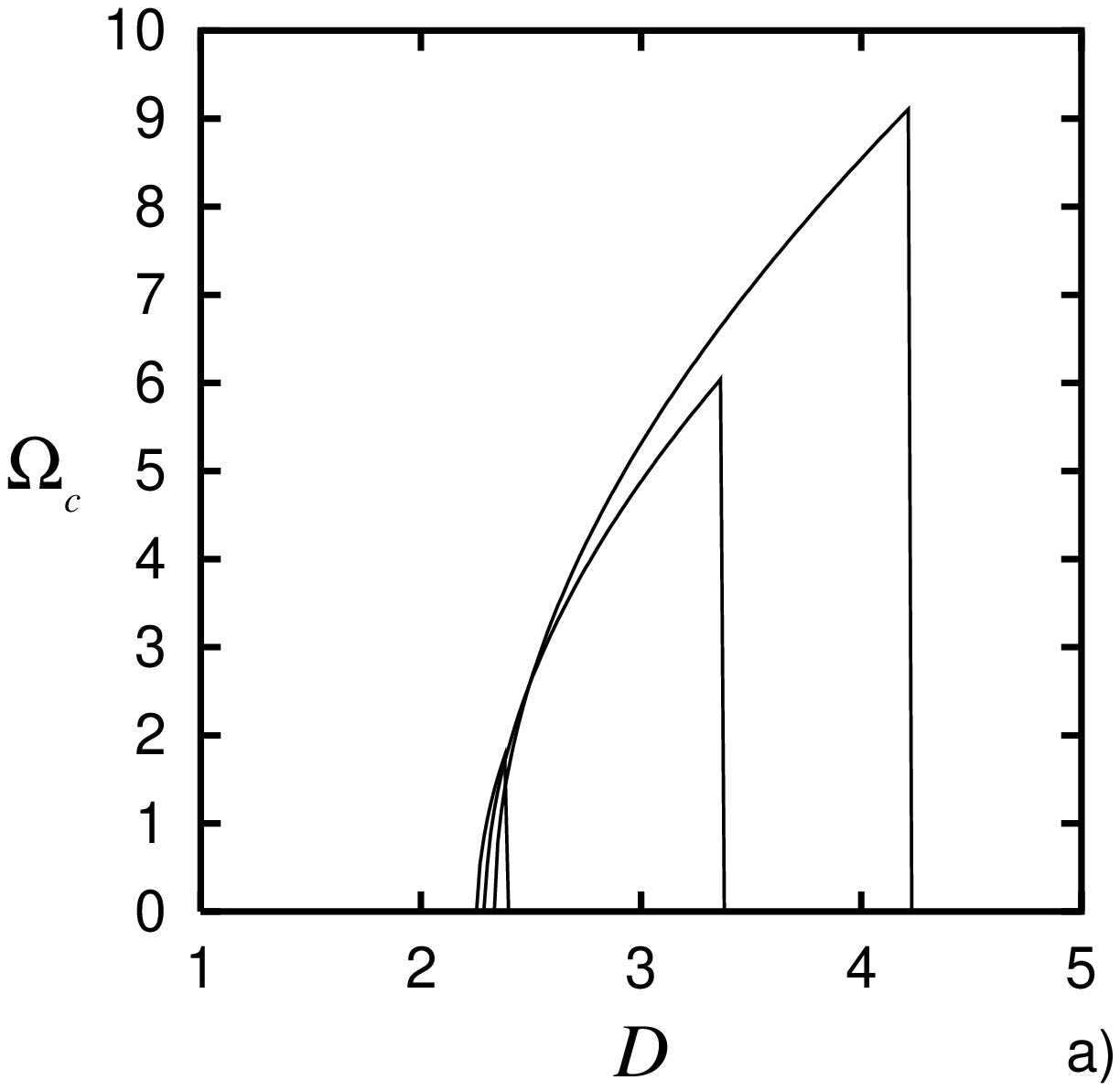, width=\linewidth}}
 \parbox[b]{\linewidth}
 {\epsfig{file=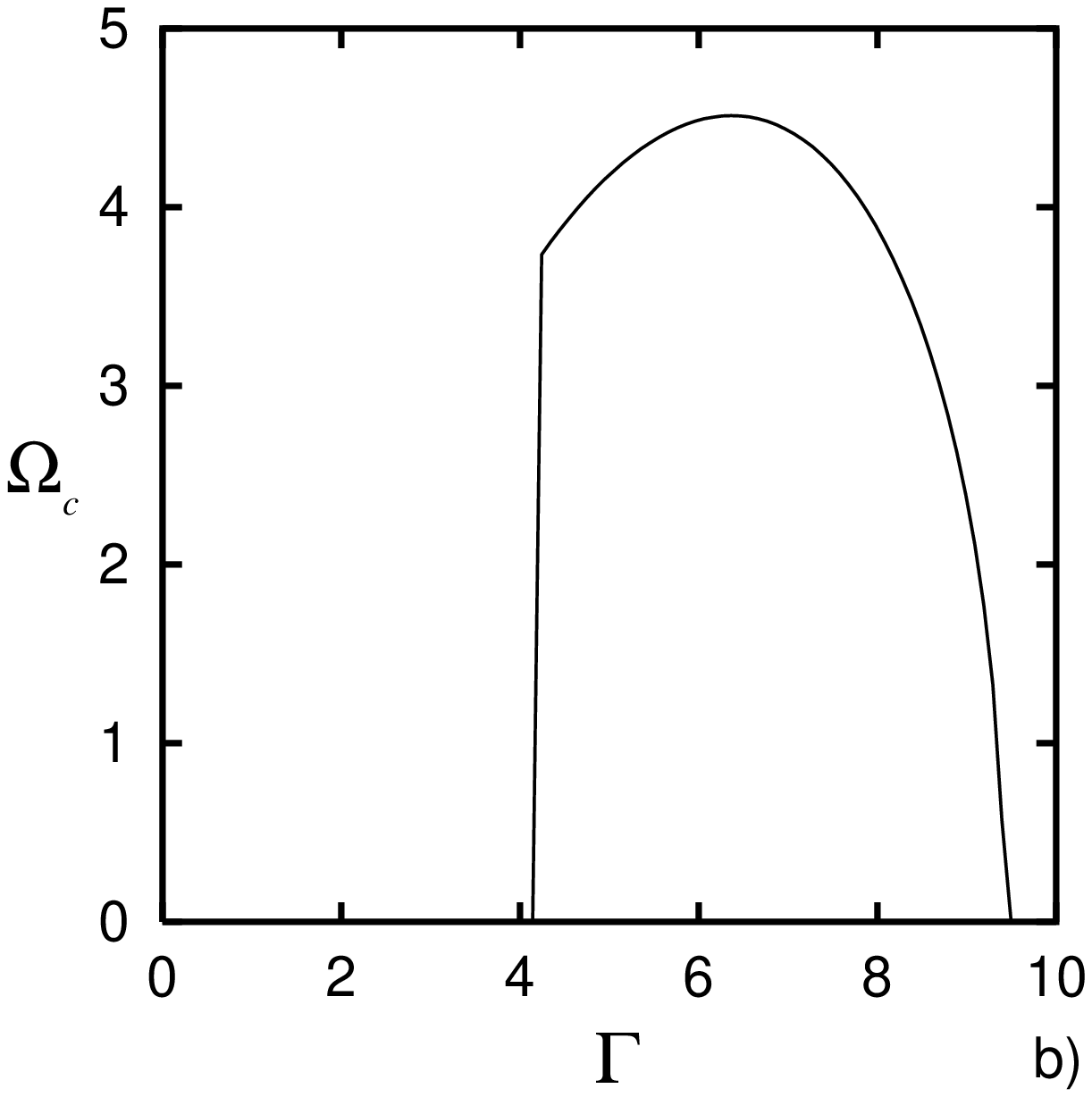, width=\linewidth}}
 \hfill\mbox{}
\end{center}
%
\caption{\captionCriticalomega}
\label{fig:criticalomega}
\end{figure}

\begin{figure}
\begin{center}
\epsfig{file=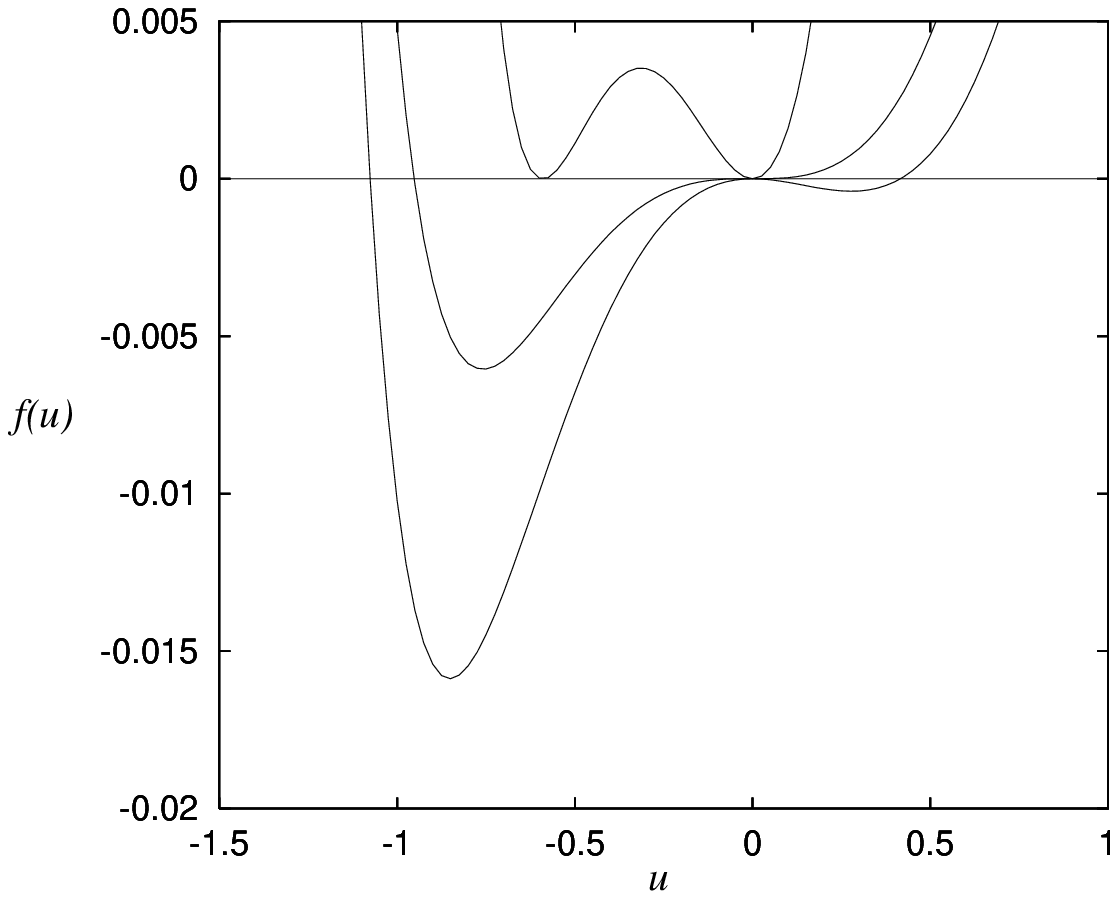, width=0.8\linewidth}
\end{center}
\caption{\captionSoftmodePotential}
\label{fig:potential}
\end{figure}

\newpage
\begin{figure}[htbp]
\begin{center}
\mbox{}
 \hfill\parbox[b]{0.45\linewidth}
{\epsfig{file=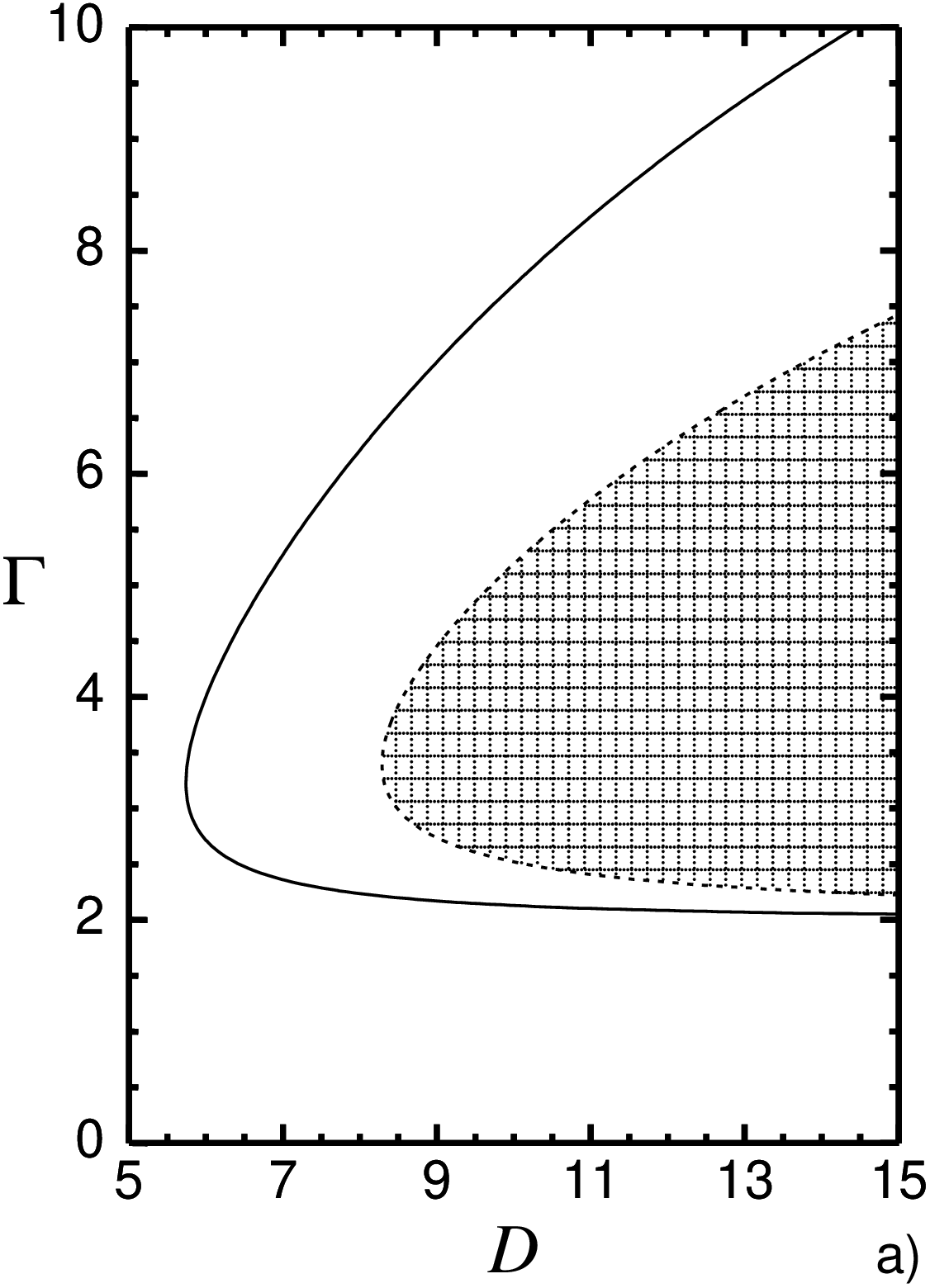, width=\linewidth}}
\parbox[b]{0.45\linewidth}
{\epsfig{file=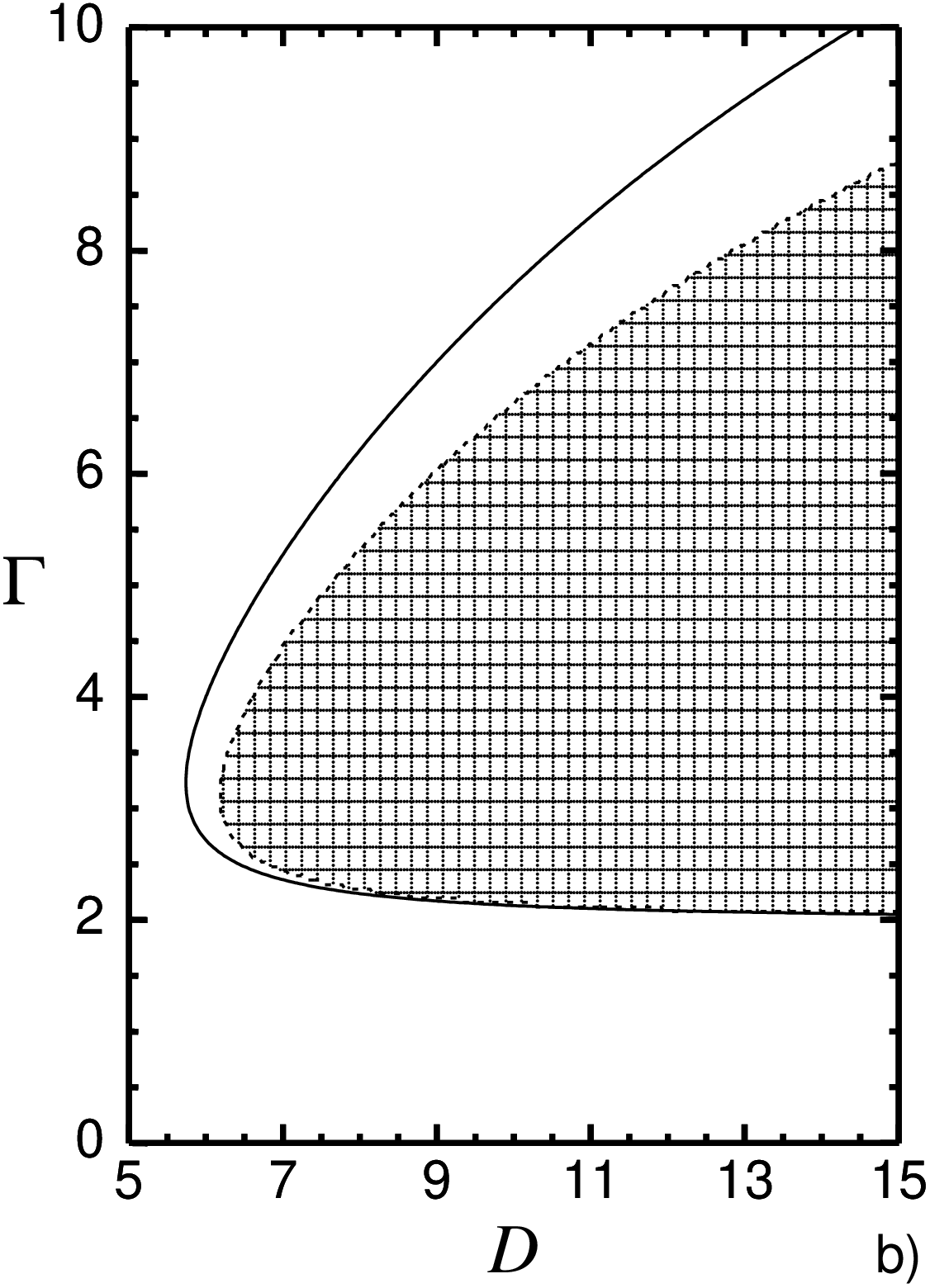, width=\linewidth}}
 \hfill\mbox{}
\end{center}
\begin{center}
 \mbox{}
 \hfill\parbox[b]{0.45\linewidth}
 {\epsfig{file=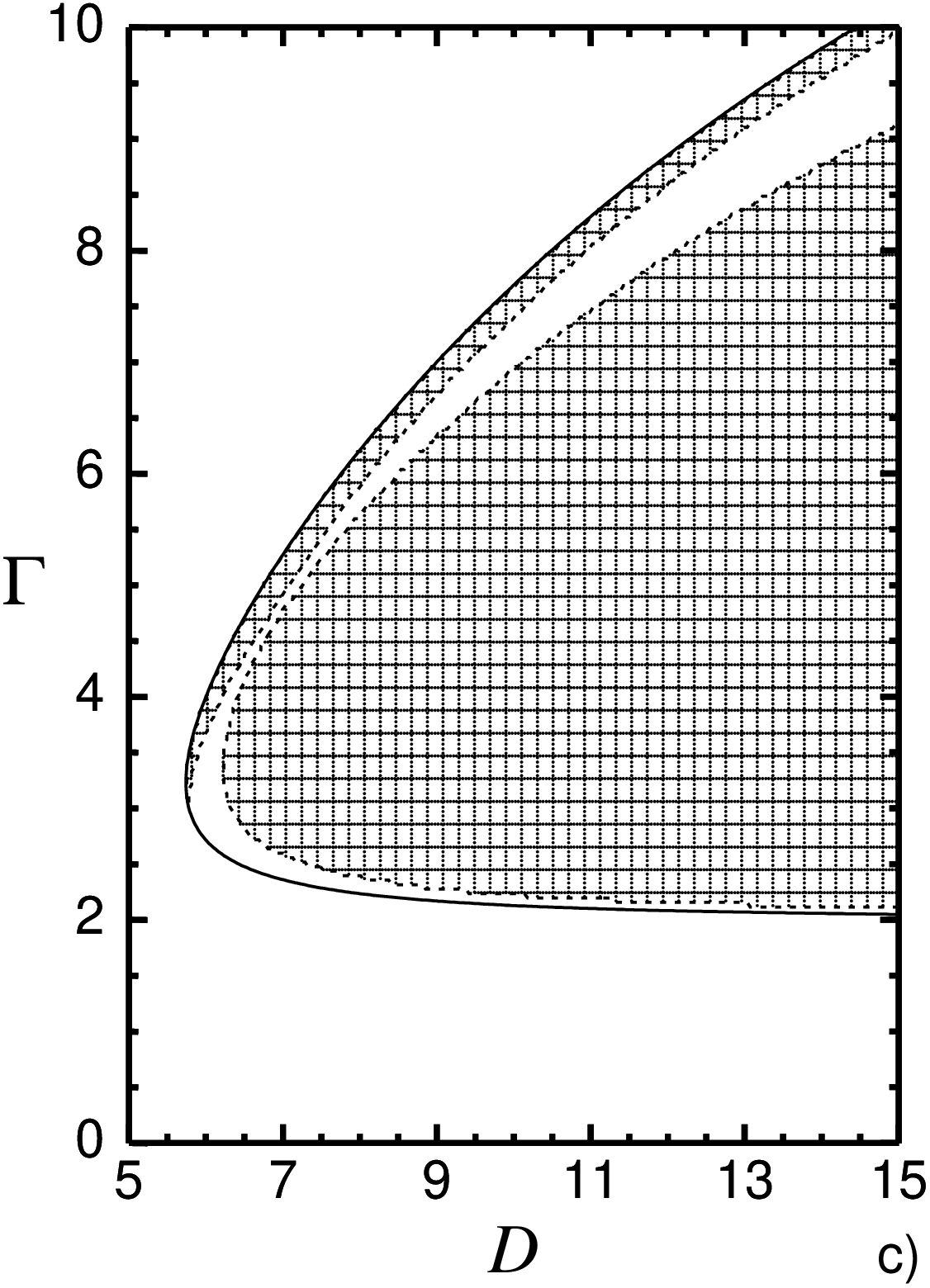, width=\linewidth}}
\parbox[b]{0.45\linewidth}
{ \epsfig{file=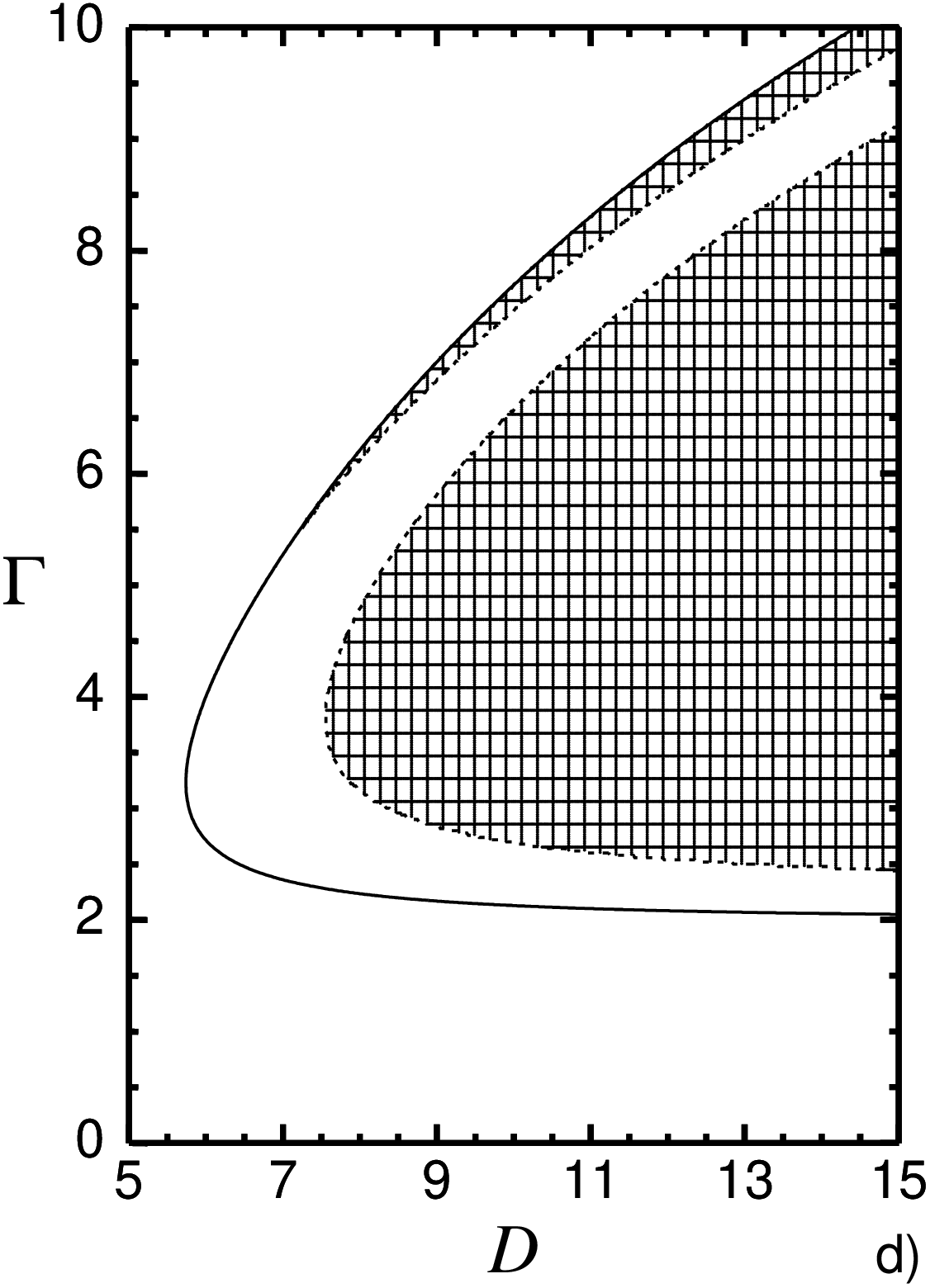, width=\linewidth}}
 \hfill\mbox{}
 \caption{\captionJDG}
 \end{center}
\label{fig:J}
\end{figure}

\end{document}